\documentclass[12pt]{article}
\usepackage{graphicx}

\newcommand\pubnumber{CIPANP2018-Benmokhtar}
\newcommand\pubdate{\today}

\def\Duquesne{Department of Physics\\
	Duquesne University, Pittsburgh, PA, 15282. USA}
\def\support{\footnote{ Supported by the U.S. National Science Foundation under grant 1615067}}
\def\Title#1{\begin{center} {\Large #1 } \end{center}}
\def\Author#1{\begin{center}{ \sc #1} \end{center}}
\def\Address#1{\begin{center}{ \it #1} \end{center}}

\newcommand\pubblock{\rightline{\begin{tabular}{l} \pubnumber\\
			\pubdate  \end{tabular}}}
\newenvironment{Abstract}{\begin{quotation}  }{\end{quotation}}
\newenvironment{Presented}{\begin{quotation} \begin{center}
			PRESENTED AT\end{center}\bigskip
		\begin{center}\begin{large}}{\end{large}\end{center} \end{quotation}}

\def\beq{\begin{equation}}
\def\eeq#1{\label{#1}\end{equation}}
\def\eeqn{\end{equation}}

\def\beqa{\begin{eqnarray}}
\def\eeqa#1{\label{#1}\end{eqnarray}}
\def\eeqan{\end{eqnarray}}

\let\bar=\overbar

\def\Dslash{\not{\hbox{\kern-4pt $D$}}}
\def\dslash{\not{\hbox{\kern-2pt $\del$}}}

\def\msb{{\bar{\ssstyle M \kern -1pt S}}}

\begin{document}
	\begin{titlepage}
		\pubblock	
		\vfill
		\Title{Probing the Strange Sea Quarks with Kaon SIDIS}
		\vfill
		\Author{Fatiha Benmokhtar\support}
		\Author{Mireille Muhoza and Collin McCauley}
		\Address{\Duquesne}
		\vfill
		\begin{Abstract}
			It is well known that protons and neutrons are made from constituents, called quarks and gluons, which give substructure to these particles. The goal of this project is to make measurements of the spatial distributions and the momenta of the quarks that provide a three-dimensional map of quarks in the nuclear medium. This knowledge provides the basis of our understanding of nuclear matter in terms of the dynamics of their internal constituents. This abstract focuses on the study of the contribution of the sea quarks and in particular of the strange sea to the proton spin structure. This study is feasible with semi-inclusive deep inelastic scattering of electrons off proton and deuteron targets in Hall B at Jefferson Lab. To achieve the desired precision, a Ring Imaging CHerenkov (RICH) detector was built so pion, kaon and proton identification is well performed in the momentum range of 3 to 8 GeV/c. The experimental method and projected precision of the measurements of the parton distributions using Kaon SIDIS will be discussed and the status of the recently built Hybrid RICH detector will be presented.
			
		\end{Abstract}
		\vfill
		\begin{Presented}
			The Conference on the Intersections of Particle and Nuclear Physics\\
			Palm Springs, CA,  May 29 - June 3, 2018
		\end{Presented}
		\vfill
	\end{titlepage}
	\def\thefootnote{\fnsymbol{footnote}}
	\setcounter{footnote}{0}

	\section{Introduction}

	Since the concept of strangeness was introduced decades ago, interest in this degree of freedom has grown exponentially. Measurements with identified strange hadrons can provide important information on several hot topics in hadronic physics: the nucleon tomography and quark orbital momentum, accessible through the study of the Generalized Parton Distribution (GPDs) and the Transverse Momentum dependent parton Distribution functions (TMDs), the quark hadronization in the nuclear medium, the hadron spectroscopy,  the search for exotic mesons and the topic of this paper: the strange helicity distribution and fragmentation functions. \\

	The understanding of the spin structure of the nucleon in terms of quarks and gluons has been the goal of intense investigations during the last decades. Results from Inclusive Deep-Inelastic Scattering (DIS) experiments, where only the scattered electron is detected, indicate that the net contribution of the quark spins to the spin of the nucleon is very small. Under the assumption of SU(3) symmetry this small contribution implies a significant negative value for the polarization of the strange quark sea in the proton. Such a value would explain the violation of the Ellis-Jaffe sum rule in inclusive DIS. However, results from HERMES for a flavor decomposition of quark helicity distributions based on Semi-Inclusive DIS (SIDIS) measurements suggest that the strange sea polarization is zero or slightly positive \cite{HERMES1, HERMES3}. This discrepancy between inclusive and semi-inclusive measurements needs to be checked by performing an independent third high precision measurement. \\  
	
	
	\section{Probing the Sea with SIDIS}

	A Jefferson Lab HallB experimental proposal \cite{prop1} was developed to further elucidate the flavor contribution to the nucleon spin (E12-09-007) through ``The study of parton distribution functions using semi-inclusive production of Kaons''.  This  A$^-$ rated experiment is granted 110 full running days using CLAS12 in Hall B of Jefferson Lab.  Hall B  will receive polarized beams of maximum energy 11~GeV and luminosity up to {$10^{35}$cm$^{-2}$s$^{-1}$}, providing a world-leading facility for the study of electron–nucleon scattering with a wide angular coverage. \\
	
\begin{figure}[!ht]
		\begin{center}
			\includegraphics[width=0.4\textwidth, angle=0]{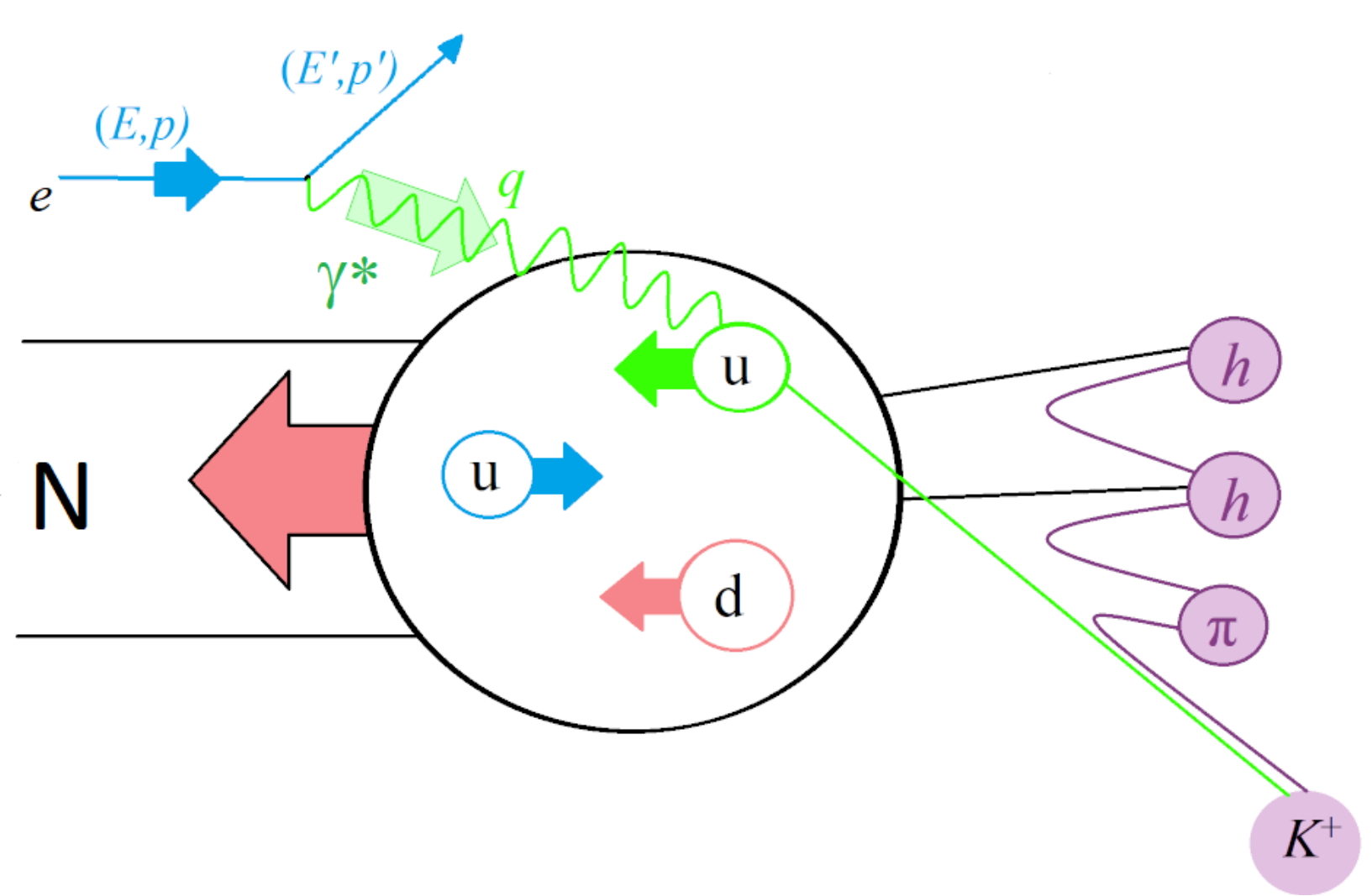}
			\caption{Hadron creation in SIDIS reaction.}
			\label{SIDIS}
		\end{center} 
	
	\end{figure}

	The semi-inclusive experiment will scatter polarized electrons off polarized deuterium and hydrogen targets. Both the scattered electron and recoiling hadron will be detected. Hadron production in DIS is described by the absorption of a virtual photon by a point-like quark followed by the fragmentation of this quark into hadronic final states, see Fig~\ref{SIDIS}. 
	
	The first process is characterized by the quark distribution functions and the second process is characterized by fragmentation functions. Fragmentation functions describe the phenomenon of hadronization, i.e., how quarks or gluons transform into hadrons before ever being detected as free particles. They are essential manifestations of Quantum Chromodynamics (QCD) and confinement. The ensuing dramatic improvement in the knowledge of fragmentation functions will be an important ingredient in the global analysis of the nucleon structure from semi-inclusive Deep-Inelastic-Scattering and pp collision data, thus contributing in a unique way to the challenge of understanding how partons build up all intrinsic properties of the nucleon, in particular how they share the nucleon's spin. This is one of the main goals of the upgraded 12 GeV CEBAF at JLab and the future Electron Ion Collider (EIC).\\
	
	\subsection{Proposed measurements}
	The proposed measurements are two-fold. First, the multiplicities for several hadron species ($\pi^+$, $\pi^-$, $\pi^0$, $K^+$ , $K^-$ , $K^s$) using both hydrogen and deuterium targets. The goal of these measurements is the control of the fragmentation functions used in the extraction of the individual quark and anti-quark contributions to the nucleon spin. In addition to the measurement of the shape ({$\it x$} dependence) of the strange parton distribution function for several {$\it z$} and {$\it {Q^2}$} bins with three independent measurements ($\pi^+$+$\pi^-$, $\pi^0$, $K^+$+$K^-$ and $K^s$).\\
	
	The second goal is the polarized measurements. The aim is to use two different methods to access the quark polarization. The first is the so called isoscalar method where only polarized deuterium is used to extract the non strange and strange polarized parton distribution functions. The second method is a full flavor decomposition method using the information on both hydrogen and deuterium targets to extract individual contributions of the quarks to the nucleon spin. Both polarized and unpolarized measurements will cover the {$\it x$} range from 0.05$~<~x~<~$0.7. 
	
	\subsubsection{The Isoscalar method}
Following the parton model, at leading order (LO), the semi-inclusive double-spin asymmetry for 
charged-kaon production can be written in terms of quark helicity distributions $(\Delta q(x,Q^2))$ 
and kaon fragmentation functions $(D_q^K(z,Q^2))$ as 
\begin{equation}
\label{eq:A-SINC}
A_1^{K}(x,Q^2\!,z) =
\frac{\sum_q e_q^2 \, \Delta q(x,Q^2)\, D_q^K(z,Q^2)}
{\sum_{q'} e_{q'}^2 \, q'(x,Q^2)\, D_{q'}^K(z,Q^2)}. 
\end{equation}\\
For a deuteron target, assuming isospin symmetry and charge conjugation invariance and by 
integrating over $z=E_h/\nu$, Eq. (\ref{eq:A-SINC}) becomes 
\begin{equation}
\label{eq:A-SINC-D}
A_{1,d}^{K}(x) =
\frac{\Delta Q(x)\,\int{{\cal D}_{Q}^K(z)dz} + \Delta S(x)\,\int{{\cal D}_{S}^K(z)dz}}
{Q(x)\,\int{{\cal D}_{Q}^K(z)dz} + S(x)\,\int{{\cal D}_{S}^K(z)dz}}
\end{equation}\\
where $Q(x)=u(x)+\bar{u}(x)+d(x)+\bar{d}(x)$, $S(x)=s(x)+\bar{s}(x)$, 
$\int{{\cal D}_{Q}^K(z)dz}= 4\int{{D}_{u}^K(z)dz} + \int{{D}_{d}^K(z)dz}$ and 
$\int{{\cal D}_{S}^K(z)dz}= 2\int{{D}_{s}^K(z)dz}$. The $Q^2$ dependence is omitted for simplicity. 
The corresponding inclusive double spin asymmetry is given by:\\ 
\begin{equation}
\label{eq:A-INC-D}
A_{1,d}(x) =
\frac{5\Delta Q(x) + 2\Delta S(x)}{Q(x) + S(x)}
\end{equation}\\
The non-strange $(\Delta Q(x))$ and the strange $(\Delta S(x))$ helicity distributions are obtained 
by combining Eqs. (\ref{eq:A-SINC-D}) and (\ref{eq:A-INC-D}) in a matrix form: \\
\begin{equation}
\label{Asym-Matrix}
\displaystyle
\left(\begin{array}{c}
A_{1,d}(x)\\[.1in]
A_{1,d}^{K}(x)
\end{array}\right)
~=~ \left(\begin{array}{c c}
P_{Q}(x) & P_{S}(x)\\[.1in]
P_{Q}^{K}(x) & P_{S}^{K}(x)
\end{array}\right)
\left(\begin{array}{c}
\Delta Q(x)/Q(x)\\[.1in]
\Delta S(x)/S(x)
\end{array}\right)
\end{equation}\\
\begin{equation}
\label{Pur_INC}
\displaystyle
{\mbox{where}}~~~~~~~~~~P_Q(x)=\frac{5Q(x)}{5Q(x) + 2S(x)},
~~~~~P_Q^K(x)=\frac{Q(x)\int{\cal D}_{Q}^K(z)dz}
{Q(x)\int{\cal D}_{Q}^K(z)dz + S(x)\int{\cal D}_{S}^K(z)dz}, 
\end{equation}
\begin{equation}\\
\label{Pur_SINC}
{\mbox{and}}~~~~~~~~~~~~P_S(x)=\frac{2S(x)}{5Q(x) + 2S(x)},
~~~~~P_S^K(x)=\frac{S(x)\int{\cal D}_{S}^K(z)dz}
{Q(x)\int{\cal D}_{Q}^K(z)dz + S(x)\int{\cal D}_{S}^K(z)dz}.
\end{equation}\\
In order to get the non-strange and the strange helicity distributions, we need to determine 
$Q(x)$, $S(x)$, $\int{\cal D}_{Q}^K(z)dz$ and $\int{\cal D}_{S}^K(z)dz$. $Q(x)$ will be taken from  the most up to date CTEQ parametrization at the time of the analysis  and $\int{\cal D}_{S}^K(z)dz$ is taken from the 
analysis of the fragmentation functions. $S(x)$ and 
$\int{\cal D}_{Q}^K(z)dz$ are to be extracted from the charged-kaon multiplicities which, for a deuteron target and at LO, are expressed by:\\ 

\begin{equation}
\label{eq:multip1}
\frac{d^2N^K(x)/dxdQ^2}{d^2N^{DIS}(x)/dxdQ^2} =
\frac{Q(x){\int{\cal D}^K_{Q}(z)dz}+S(x){\int{\cal D}^K_{S}(z)dz}}{5Q(x)+2S(x)}.
\end{equation}\\

By neglecting the term 2S(x) compared to 5Q(x) we get 
\begin{equation}
\label{eq:multip2}
S(x)\int{\cal D}^K_{S}(z)dz = Q(x)\left[5\frac{d^2N^K(x)}{d^2N^{DIS}(x)} - 
\int{\cal D}^K_{Q}(z)dz \right].
\end{equation}

	\subsubsection{Sensitivity to {$\Delta{\bar{u}}-\Delta{\bar{d}}$}} 
	Measuring the spin asymmetry for $K^-$ jet off a hydrogen target is of a special interest since it is particularly sensitive to $\Delta$${\bar{u}}$-$\Delta$${\bar{d}}$  which provides an excellent test for theoretical models describing the unpolarized densities ${\bar{u}}-{\bar{d}}$. Two of such models are the Chiral Quark Soliton Model (${\aleph}$QCM) \cite{Dress}, which is based on an effective theory where baryons appear as soliton solutions of the Chiral Lagrangian. The second one is a meson cloud model \cite{Fu-Guang} that describes the nucleon as bare nucleon surrounded by a cloud of virtual mesons. 
	\begin{figure}[ht]
		\begin{center}
			\includegraphics[width=1.\textwidth, angle=0]{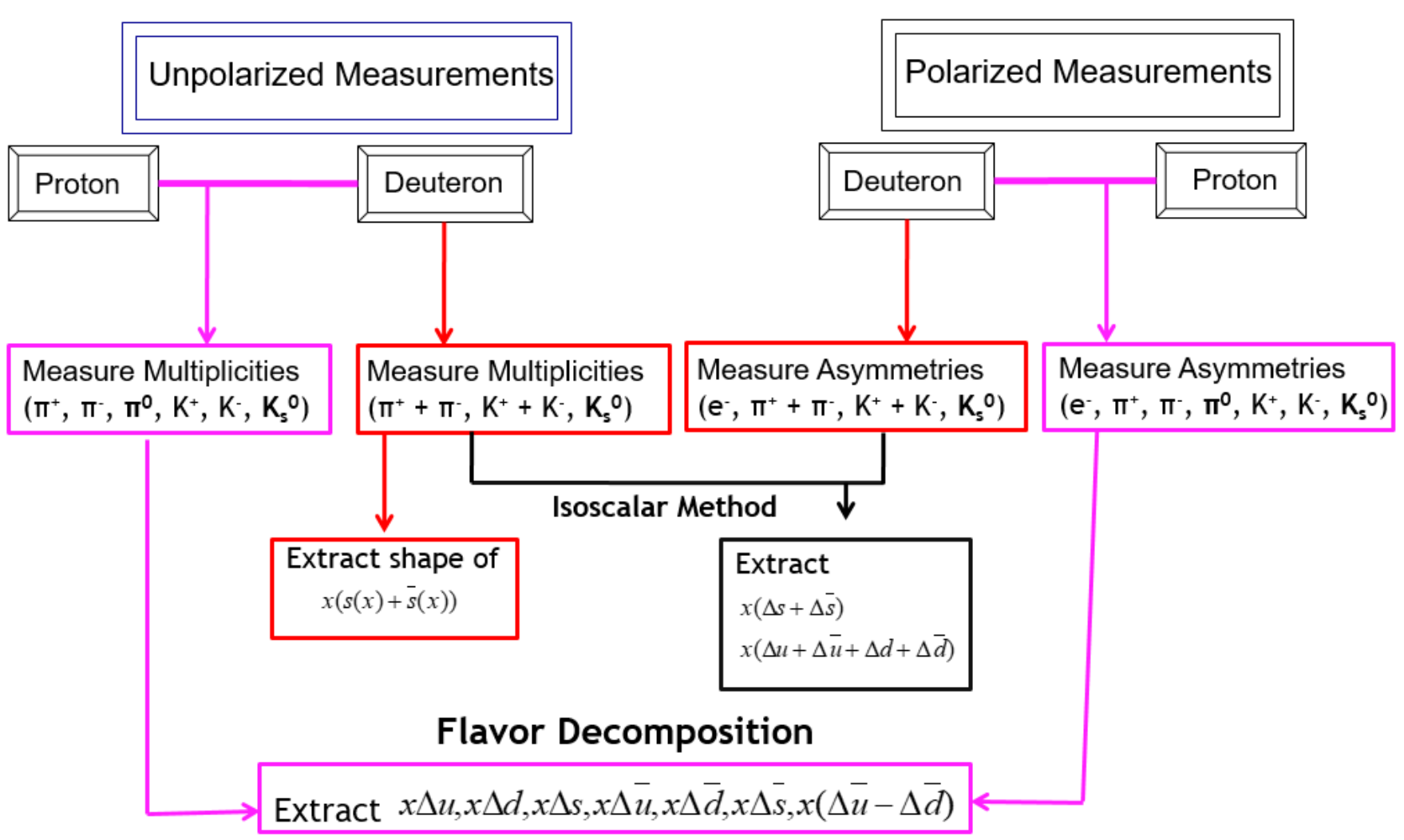}
			\caption{Program proposed by the E12-09-007 Experiment in CLAS12 at Jefferson Lab. }
			\label{Program}
		\end{center} 
	\end{figure}	
	To achieve the proposed measurements, a Ring Imaging detector is considered as an important addition to the CLAS12 instrumentation to make a clean particle identification. 
	
	\subsection{The CLAS12 RICH Detector} 
	A Ring-Imaging CHerenkov (RICH) detector is a device that allows the identification of electrically charged subatomic particle types through the detection of the Cherenkov radiation emitted (as photons) by the particle in traversing a medium with refractive index $ n > 1$ . The identification is achieved by measurement of the angle of emission, $\theta_c$ , of the Cherenkov radiation, which is related to the charged particle's velocity {\it v} by $\cos \theta_c = \frac{c}{nv}$  where {$\it{c}$} is the speed of light. Particles of different masses but same momentum can be distinguished by the distinct contours of their emitted photons. Photons emitted by heavier particles (like Kaons)  fall in smaller radius circle than the ones emitted by a lighter particles (like Pions).   \\

		In addition to the E12-09-007 experiment; the PAC of Jefferson Lab  approved two other proposals aiming to study kaons with with CLAS12 as well: Studies of Boer-Mulders effect with kaons (E12-09-008) and Studies of Kotzinian-Mulders effect with kaons (E12-09-009). All three experiments demand an efficient hadron identification across the entire momentum range from 3 to {8~GeV/$c$} and scattering angles up to {$25{^o}$}. A pion rejection factor of about {$1:500$} is required to limit the pion contamination in the kaon sample to a few percent level.\\
		
\begin{figure}[!ht]
				\includegraphics[width=.45\textwidth, angle=0]{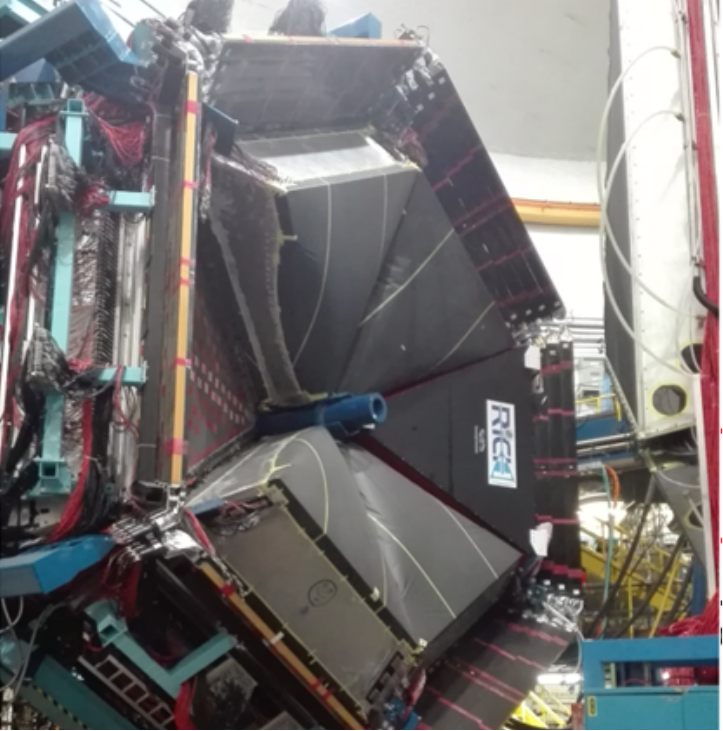}
					\includegraphics[width=.5\textwidth, angle=0]{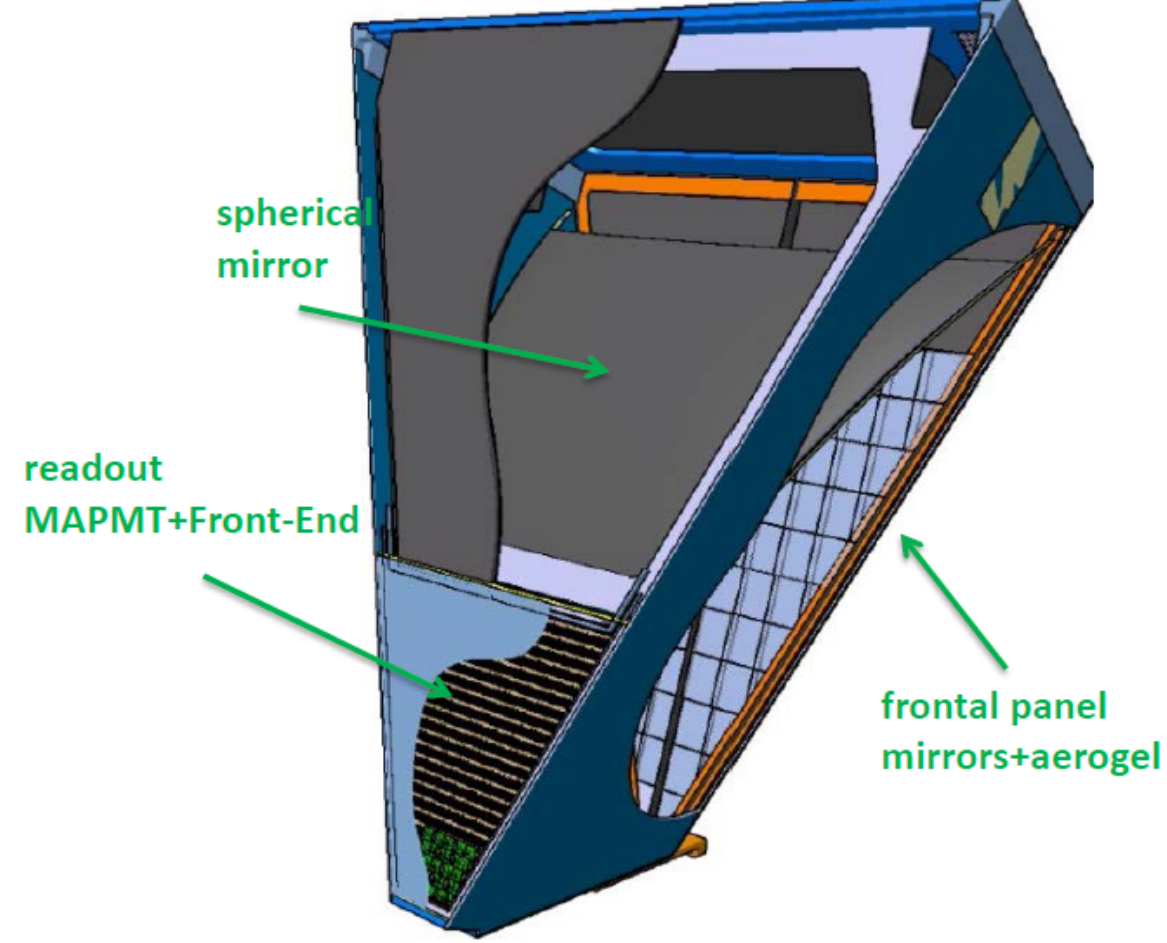}
				\caption{Left: RICH installed in CLAS12. Right: RICH components.}
				\label{cerdet}
				
		\end{figure}

		The CLAS12 baseline equipment comprises a time-of-flight system(TOF), able to efficiently identify hadrons up to a momentum of about {3~GeV/$c$}, and two Cherenkov gas detectors of high (HTCC) and low (LTCC) threshold, reaching the needed pion rejection power only close to the upper limit (around {7~GeV/${c}$}) of hadron momenta and are not able to distinguish kaons from protons.  A ring-imaging Cherenkov detector (RICH) has been proposed by the Italian INFN groups of Bari, Ferrara, Laboratori Nazionali di Frascati and ISS$/$Roma1, together with the US Jefferson Lab, Duquesne University, Argonne National Lab, University of Glasgow. These different institutions formed the ``CLAS12 RICH group''.  One RICH detector is constructed and installed in CLAS12 at the end of 2017. Another RICH is under construction. These two RICH detectors will replace two symmetric LTCC radial sectors out of a total of six. Fig.~\ref{cerdet} shows the one RICH detector installed in CLAS12. This detector is a hybrid imaging RICH design incorporating Aerogel radiators, Hamamatsu Multi-Anode Photo-Multipliers (MAPMTs) for visible light photon detection, and a focusing mirror system including one spherical many planar mirrors. 
		
\subsection{Expected Precisions} 
\begin{figure}[!ht]
	 	\begin{center}
	 		\includegraphics[width=1.05 \textwidth, angle=0]{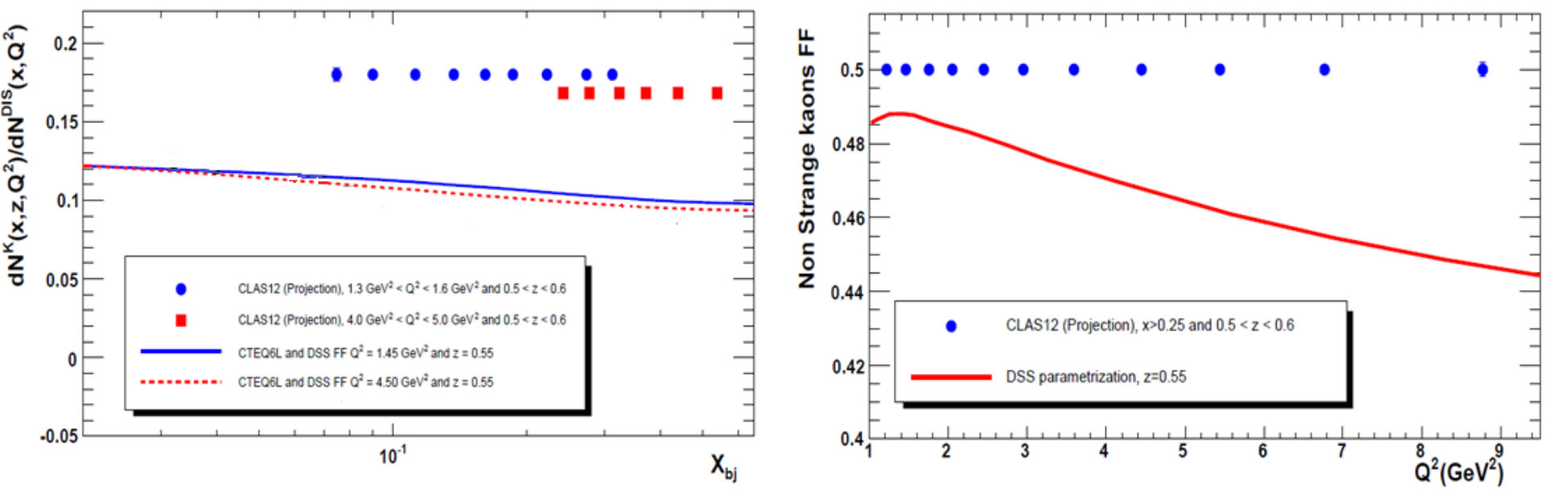}
	 		\caption{Prediction of kaon multiplcities (left) and non strange fragmentation functions (right) for the proposed experiment.}
	 		\label{mult}
	 	\end{center} 
 	\end{figure}

    Fig~\ref{mult} summarizes the projected kaon multiplicities (left) and non strange fragmentation functions (right) from the E12-09-007 experiment. 
    
\begin{figure}
		\begin{center}
				\includegraphics[width=0.55\textwidth,angle=0]{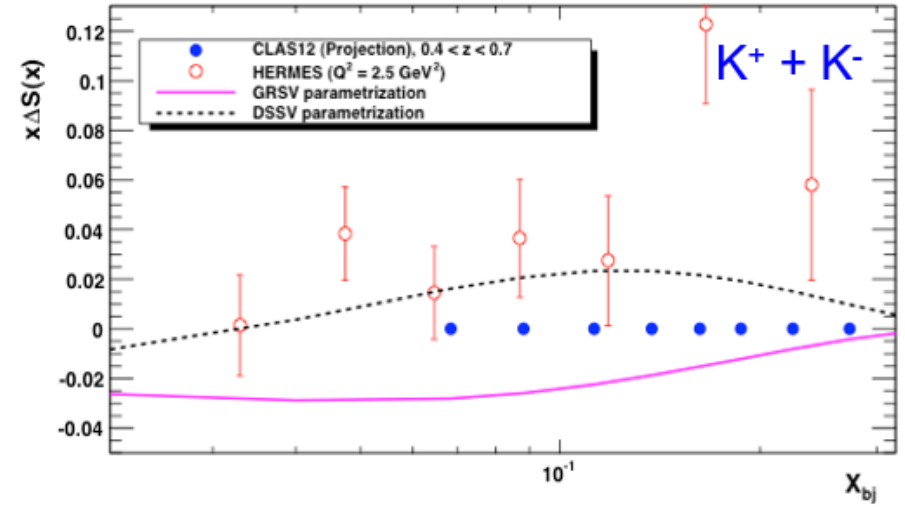}
				\includegraphics[width=0.55\textwidth,angle=0]{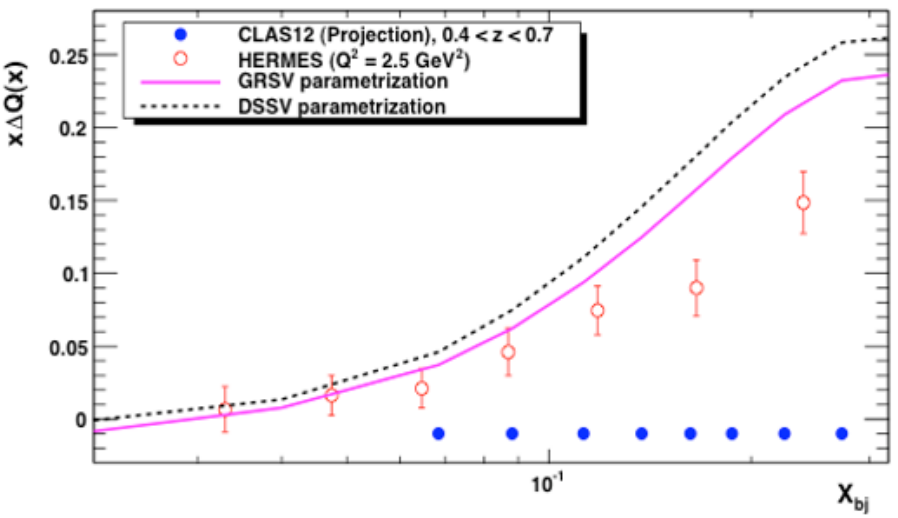}
				\caption{Statistical projections for the Isoscalar method measurements. Top figure:  $x\Delta{S(x)}$ and bottom figure: $x\Delta{Q(x)}$}
		\label{xdeltax}
		\end{center}
	\end{figure}

	Fig.~\ref{xdeltax} summarizes the projected precision form the future  E12-09-007 isoscalar measurements in comparison to {$GRSV$} \cite{grsv} and {$DSSV$} \cite{dssv} parameterizations along with the previously cited HERMES data.

\begin{figure}
       	\begin{center}
		\includegraphics[width=0.6\textwidth, angle=0]{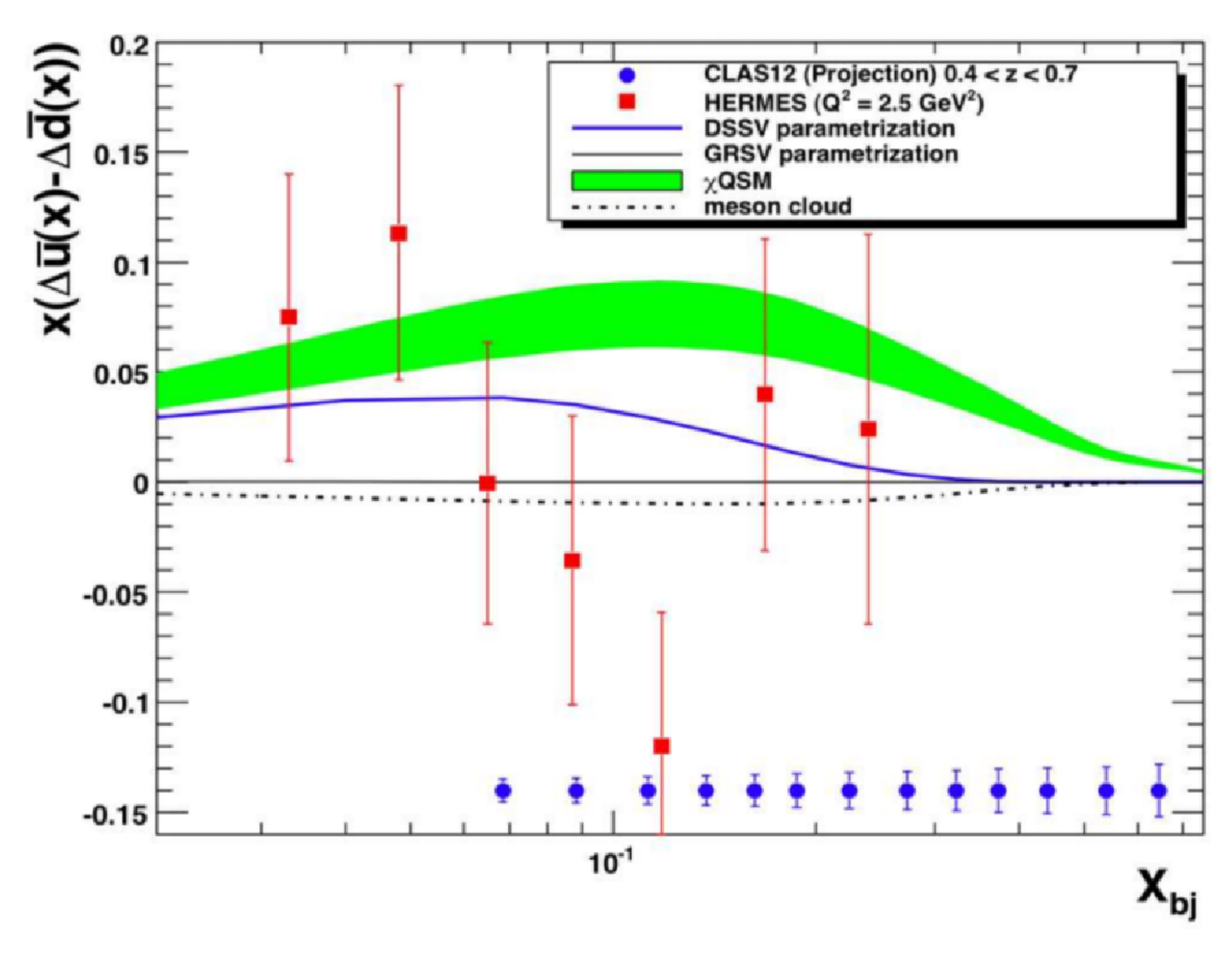}
		\caption{Statistical projections for $x(\Delta{\bar{u}}-\Delta{\bar{d}})$}
		\label{xdeltaxbar}
	\end{center}
\end{figure}

	 Fig.~\ref{xdeltaxbar} shows the existing data on ${\bar{u}}-{\bar{d}}$ measured at HERMES. While the data tend to prefer flavor symmetry, no decisive conclusion can be made due to the lack of statistical precision. The blue points in the figure are the projected high statistics data one can achieve with CLAS12. \\

\section{Outlook}
 CLAS12 at Jefferson Lab underwent recently an upgrade to 12 GeV electron beam energies. The proposed experiment will run in 2020 after a second RICH sector is built and after a polarized deuterium target is build as well. Calibrations of the first RICH detector are underway and first analysis show that the detector is healthy. We expect final analysis of the kaon SIDIS data to be achieved a year or two after it is collection. 


\begin{thebibliography}{}
	\bibitem{NIMpaper}
	``The large-area hybrid-optics CLAS12 RICH detector: Tests of innovative components'':  M. Contalbrigo $et al.$ Nuclear Instruments \&  Methods in Physics Research A (2014), ${http://dx.doi.org/10.1016/}$ j.nima.2014.06.072i
	
	\bibitem{witchger}
	 ``Investigation of Hamamatsu H8500 phototubes as single photon detectors" M. Hoek, V. Lucherini, M. Mirazita, R.A. Montgomery, A. Orlandi, S. Anefalos Pereira,S. Pisano, P. Rossi, A. Viticchi ,  {\bf A. Witchger} arXiv:1409.3622 [physics.ins-det].  
	
	\bibitem{prop1}
	  "Studies of Partonic Distributions using  Semi-Inclusive Production of  Kaons" {\bf{F. Benmokhtar}}, H. Hafidi, A. El-Alaoui and M. Mirazita: Jefferson Lab CLAS12-PR-09-007 Experimental proposal.
	
	\bibitem{HERMES3}
	 A. Airapetian {\it et al.} Phys. Lett. B{\bf{666}}446 2008  
	\bibitem{HERMES1} A. Airapetian {\it et al.}, Phys. Rev. D {\bf{71}}  2003, 2005
	\bibitem{HERMES2}
	 A. Airapetian {\it et al.}, Phys. Rev. Lett {\textbf{94}}, 012002, 2005
	\bibitem{krex}
	 S. Kretzer, Phys. Rev. D {\textbf{62}}, 054001, 2000
	\bibitem{Dress}
	 B.~Dressler, K~Goeke, M.~V.~Polyakov and C~Weiss, EPJ,{\bf{C14}},147, 2000.
	\bibitem{Fu-Guang}
	 F.~Cao and A.~I.~Signal, Phys. Rev. D {\textbf{68}}, 074002,2003
	\bibitem{grsv}M. Gluck {\it et al.}, Phys. Rev. D {\bf{63}},094005, 2001
	\bibitem{dssv}
	 D. deFlorian {\it et al.} , Phys. Rev. Lett. {\bf{101}}, 072001, 2008
	
	
\end{thebibliography}


					
		
	\end{document}